\documentclass[fleqn,twoside]{article}
\usepackage{espcrc2}

\usepackage{graphicx}
\usepackage[figuresright]{rotating}

\newcommand\be{\begin{equation}}
\newcommand\ee{\end{equation}}
\newcommand\bea{\begin{eqnarray}}
\newcommand\eea{\end{eqnarray}}
\newcommand{\bdm}{\begin{displaymath}}
\newcommand{\edm}{\end{displaymath}}
\newcommand\nn{ \nonumber\\}
\newcommand\n{ \nonumber}

\newcommand{\<}{\langle}
\renewcommand{\>}{\rangle}

\newcommand{\AmS}{{\protect\the\textfont2
  A\kern-.1667em\lower.5ex\hbox{M}\kern-.125emS}}

\title{M\"obius Fermions: Improved Domain Wall Chiral Fermions
        \thanks{This is supported in part by Department of Energy 
         under Contracts No. DE-FG0291ER40676 and DFFC02-94ER40818}
}

\author{Richard C. Brower\address[BU]{Physics Department, Boston University,
Boston, MA 02215, USA },
Hartmut Neff\addressmark[BU]
and Kostas Orginos\address[MIT]{Center for Theoretical Physics,
MIT, Cambridge , MA 02139, USA}}
       
\begin{document}

\begin{abstract}
  A new class of domain wall fermions is defined that interpolates between
  Shamir's and Bori\c{c}i's form without increasing the number of Dirac
  applications per CG iteration.  This class represents a full (real)
  M\"obius transformation of the Wilson kernel.  Simulations on quenched
  Wilson lattices with $\beta = 6.0$ show that the number of lattice sites
  ($L_s$) in the fifth dimension can be reduced by a factor of 2 or more at
  fixed value of chiral symmetry violations measured by the residual mass 
  ($m_{res}$).
\vspace{1pc}
\end{abstract}

\maketitle

\section{INTRODUCTION}

Chiral fermions based on the Ginsparg-Wilson relations give a rigorous
approach to exact chiral symmetry at finite lattice spacing. However this
clear theoretical advance requires efficient algorithms to allow them to be
used routinely in lattice gauge theory simulations.  The two major
implementations base on the overlap form of Neuberger~\cite{Neuberger:1998fp} or the domain wall of Kaplan~\cite{Kaplan:1992bt},
adapted by Shamir~\cite{Shamir:1993zy} and Bori\c{c}i~\cite{Borici:2004pn} can both be
seen as a means to satisfy the zero quark mass Ginsparg-Wilson relation,
$$
\gamma_5  D_{ov}(0) +   D_{ov}(0) \gamma_5  - 2 D_{ov}(0) \gamma_5 D_{ov}(0)
= 2 \gamma_5 \Delta_{L_s}  ,\n
$$
where the error $\Delta_{L_s} \rightarrow 0$ is made as small as needed by
taking $L_s \rightarrow \infty$, where $L_s$ parameterizes the order of a
rational polynomial approximation.  A general solution, including a non-zero
quark mass m, is provided by the  overlap operator,
\be
D_{ov}(m) = \frac{1+m}{2} + \frac{1-m}{2}  \gamma_5 \epsilon_{L_s}[\gamma_5
D(M_5)] \; ,
\ee
where $\epsilon_{L_s}[x]$ is the rational approximation of the sign function,
$ \epsilon[x] = x/|x| $ and $\Delta_{L_s}[x] \equiv (1 -
\epsilon^2_{L_s}[x])/4$.  In fact the GW relation defines an infinite class
of chiral fermions depending on the selection of the kernel $D(M_5)$, the
simplest but certainly not the ideal example being the Wilson fermion
operator itself,
$$
\begin{array}{l}
D_w(M_5) =  (4+M_5) \delta_{x,y} - \cr
 \frac{1}{2} \Bigl[  (1 - \gamma_\mu) U_\mu(x) \delta_{x+\mu,y} 
+  (1 + \gamma_\mu) U_\mu^\dagger(y) \delta_{x,y+\mu} \Bigr] \; , 
\end{array}
\label{eq:D_w} 
$$
with mass $M_5 = O(-1)$.  On the other hand the standard Shamir implementation
of Domain Wall fermions is equivalent to an overlap kernel,
\be
D_{Shamir}(M_5) = \frac{a_5 D_w(M_5)}{2 + a_5 D_w(M_5)}  \; .
\ee
Here we suggest a modification of the standard Shamir domain wall approach
that replaces the overlap kernel, or 5-th time Hamiltonian $\gamma_5 D(M_5)$, by a
real M\"obius transformation of the Wilson operator: $D_w(M_5) \rightarrow (a
+b D_w(-1)/(c + d D_w(-1))$ or equivalently in a more convenient
parameterization for the domain wall formulation,
\be
D_{Mobius}(M_5) =  \frac{(b_5 + c_5)D_w(M_5)}{2 +  (b_5 - c_5) D_w(M_5)} \; .
\ee
In addition to the Shamir parameter, $a_5 = b_5 -c_5$, there is a new
independent scaling factor, $\alpha = c_5 + b_5$, which turns out to be a significant
advantage at finite $L_s$ even though it is irrelevant at $L_s = \infty$ due
scale invariance of the sign function: $\epsilon[\alpha x]= \epsilon[x]$.
One way to understand the advantage of the M\"obius transform is to notice
that it can map any 3 real values (e.g.  eigenvalues of $D_w$) to arbitrary
points.  This allows one to choose $M_5$ to separate legitimate chiral modes from
doublers and simultaneously to adjust both the smallest and largest real
eigenvalue in $D(M_5)$.  Note also M\"obius fermions have a continuous path to
conventional overlap kernel as implemented by Bori\c{c}i's with $ b_5 = c_5 = a_5$.

\section{DOMAIN WALL IMPLEMENTATION}
\label{sec:DW}

It is very natural to implement the real M\"obius kernel by 
the 5-d Domain Wall matrix, $D_{DW}(m)$:
$$
 \left[
\begin{array}{cccc}
D^{(1)}_+ & \quad D^{(1)}_- P_- & \cdots &  -mD^{(1)}_- P_+  \\
\quad D^{(2)}_- P_+ & D^{(2)}_+ & \cdots & 0  \\
\cdots & \cdots & \cdots & \cdots   \\
-mD^{(L_s)}_- P_- & 0 &  \cdots & D^{(L_s)}_+  \\
\end{array} \right]
$$
where $D^{(i)}_+ = b_5 \omega_i D_w(M_5) +1$ and $D^{(i)}_- = c_5 \omega_i
D_w(M_5) - 1$. In this discussion we simplify to the case of  polar decomposition
($\omega_i = 1$), although Zolotarev polynomials ($\omega_i \ne 1$)  can be implemented as well.
By standard LDU decomposition~\cite{Edwards:2000qv} for $D_{DW}(m)$, this leads back to an effective
overlap operator, with
\be
\epsilon_{L_s}[H] = \frac{(1+H)^{L_s/2} - (1-H)^{-L_s/2}}{(1+H)^{L_s/2} + (1-H)^{-L_s/2}} 
\ee
given by the polar decomposition for the ``Hamiltonian'' $H = \gamma_5
D_{Mobius}(M_5)$ defined above. The detailed algebraic steps will be
published in a forthcoming article.  The chiral modes at the boundary
\bea
q_x  &=&  P_- \Psi_{x,1} + P_+\Psi_{x,L_s} \nn
\bar q_y &=&  - [\bar \Psi_1 D^{(1)}_-]_y  P_+ - [\bar \Psi_{L_s}  D^{(L_s)}_-]_y P_- \n
\eea
give the direct connection to the overlap propagator,
\be
\<q_x \bar q_y\> = \frac{1}{1-m}[D_{ov}^{-1}(m) - 1]_{xy}   \; .
\label{eq:qqbar}
\ee
A key observation for this action (as well as Bori\c{c}i's
and Chiu's examples)  is that the new version of ``gamma 5'' Hermiticity
for the M\"obius domain wall requires
pulling out a factor $D_- = Diag[D^{(1)}_-,
D^{(2)}_-, \cdots, D^{(L_s)}_-]$ as well as reflecting, ${\cal R}$, in
the fifth dimensions:
$$\gamma_5 {\cal R} D^{-1}_- D_{DW}(m) = (D^{-1}_- D_{DW}(m))^\dagger
\gamma_5 {\cal R}$$ 
This in turn is reflected in the proper definition of chiral boundary states.

\section{DEFINITION OF RESIDUAL MASS}

The chiral breaking operator $\Delta_{L_s}(x)$ defined above provides
the breaking term in the 4-d Noether theorem,
$$
\delta (\bar \psi D_{ov}(m) \psi) =  m\bar \psi ( \gamma_5 + \widehat
\gamma_5) \psi + 2 (1-m) \bar \psi \gamma_5\Delta_{L_s} \psi .
$$
Therefore its matrix elements are the proper measure
of the approach to exact chirality for $L_s \rightarrow \infty$. Physically
relevant matrix elements should be sensitive to long distance (IR) physics.
In the Domain Wall language, there is a corresponding breaking
term in axial Ward-Takahashi identity,
$$
\Delta_\mu J^{a, DW}_\mu(x)= 2 m \, \bar q_x \lambda^a \gamma_5 q_x 
+ 2
\bar Q_x \gamma_5 \lambda^a Q_x
$$
where the $Q_x, \bar Q_x$ fields lie on {\bf any} plane separating the
left and right domain walls. A convenient measure of the breaking term (or
residual mass) is given by
$$
m_{res}(t)  \equiv  \frac{\sum_{\vec x}  \< \bar Q_{\vec x,t} \gamma_5 Q_{\vec x,t} \; \bar q_0 \gamma_5 q_0  \>_c}
             {(1-m)^2 \sum_{\vec x}  \< \bar q_{\vec x,t} \gamma_5 q_{\vec x,t} \; \bar q_0 \gamma_5 q_0  \>_c } 
$$
in the plateau region with $t$ away from the sources. An alternative
definition (equivalent in infinite volumes) is found by summing over all time
slices and removing the contact term in Eq.~\ref{eq:qqbar}.  This expression
for $m_{res}$ measures a specific matrix element of $\Delta_{L_s}$,
$$
m_{res}  \equiv  \frac{Tr[ \Delta_{L_s}(H) D_{ov}^{-1} D_{ov}^{\dagger -1}]}{
   Tr[  D_{ov}^{-1} D_{ov}^{\dagger -1}]} 
  =         \sum_{\lambda} \; \rho(\lambda)\; \Delta_{L_s}(\lambda)
$$
The positive definite kernel guarantees that zero residual mass implies the
exact Ginsparg-Wilson relations and unbroken Ward-Takahashi relations.

\section{NUMERICAL RESULTS}

In order to have the performance of M\"obius (or Bori\c{c}i) fermions on a equal
footing with Shamir, one must be able to use even-odd preconditioning.  However
the 5-d even-odd partition is impractical because the new operator $D_-$
connects even and odd sites so that one must invert it in each CG step.
Instead we define the even-odd pattern only on each 4-d slice, 
not alternating in 5th axis. Now the result is that even-even and odd-odd matrices are
independent of the gauge fields and can be inverted analytically at
negligible cost. Performance tests have shown that this new 4-d even-odd
precondition results in an improvement factor of roughly 3 just as the 5-d
even-odd had done for Shamir.

\begin{figure}[htb]
%\vspace{9pt}
\includegraphics[scale=0.6]{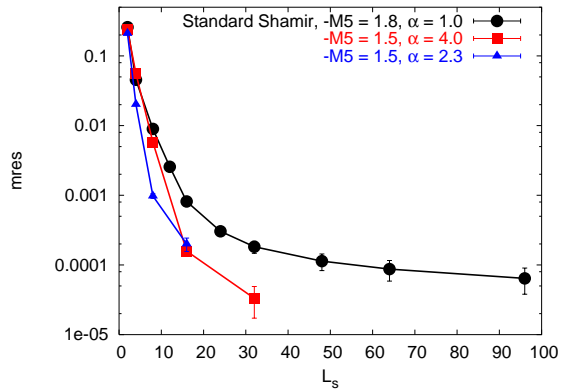}
\caption{Comparison of  $m_{res}$ vs $L_s$ for M\"obius vs standard Shamir.}
\label{fig:mobius}
\end{figure}

We tested the chirality and convergence of the M\"obius domain wall operator
on 20 quenched lattices at $\beta=6.0$ from the NERSC gauge connection. Our results were
compared with the standard Shamir fermions for a bare quark mass $m = 0.06$,
resulting in a pion mass of roughly 0.4 (in lattice units) for $L_s=16$. All
comparisons were done for standard Shamir operator with $a_5\equiv b_5 - c_5
=1.0$. We tuned $M_5$ and the scale $\alpha$.  The pion masses were adjusted
so that $L_s=8$ coincided with Shamir at $L_s=16$.  For $L_s=8$ our tests
found that $M_5=-1.5$ is the optimal choice, which is not surprising since
it sits at the mid point between the two Aoki phases at -2.2 and -0.8. We
found for $L_s = 8$ that the optimal scale factor was $\alpha = 2.3$.
Furthermore we explored $m_{res}$ at larger $L_s$. There appears to be a
crossover at $L_s = 16$ for $\alpha = 4$.

More detailed performance data for a variety of quench and unquenched
lattices is needed, but our preliminary observations are (1) at $m_{res} =
O(0.1\%)$ suitable for HMC simulations, $L_s = 8$ gives essentially the same
$m_{res}$ as conventional Shamir with $L_s = 16$, (2) at large values of
$L_s$ suitable for valence quarks, $m_{res}$ drops very rapidly if we use
large scale factors and (3) finally at fixed $m_{res}$, the number
of CG iterations increases only modestly for larger scale factors.

\end{document}